# Might Europe one day again be a global scientific powerhouse? Analysis of ERC publications suggests it will not be possible without changes in research policy

Alonso Rodríguez-Navarro[a,b], Ricardo Brito[a]

[a] *Departamento de Estructura de la Materia, Física Térmica y Electrónica y GISC, Universidad Complutense de Madrid, Plaza de las Ciencias 3, 20040, Madrid, Spain*
[b] *Depatamento de Biotecnología-Biología Vegetal, Universidad Politécnica de Madrid, Avenida Puerta de Hierro 2, 28040, Madrid, Spain*

Numerous EU documents praise the excellence of EU research without empirical evidence and against academic studies. We investigated research performance in two fields of high socioeconomic importance, advanced technology and basic medical research, in two sets of European countries, Germany, France, Italy, and Spain (GFIS), and the UK, the Netherlands, and Switzerland (UKNCH). Despite historical and geographical proximity, research performance in GFIS is much lower than in UKNCH, and well below the world average. Funding from the European Research Council (ERC) greatly improves performance both in GFIS and UKNCH, but ERC-GFIS publications are less cited than ERC-UKNCH publications. We conclude that research performance in GFIS and in other EU countries is intrinsically low even when it is generously funded. The technological and economic future of the EU depends on improving research, which requires structural changes in research policy within the EU, and in most EU countries.

**1. Introduction**

Compelling evidence from academic studies demonstrates the weakness of European research (Albarrán et al. 2010; Bauwens et al. 2011; Bonaccorsi 2007; Bonaccorsi et al. 2017a; Dosi et al. 2006; Herranz and Ruiz-Castillo 2013; Rodriguez-Navarro and Narin 2018; Rodríguez-Navarro and Brito 2018b), especially in fields that are at the forefront of technological knowledge (Bonaccorsi 2007; Rodriguez-Navarro and Narin 2018; Sachwald 2015). This situation has never been addressed by the European Commission (EC), which on the contrary has developed research policies assuming that European research is excellent. For many years, EU research policy was based around the "European paradox," which proclaims strong research and weak innovation in the EU. Recently, references to this paradox have disappeared from EC documents, but the notion is as present as it was when it was defined 24 years ago (European Commission 1995).

A report from an independent *High Level Group* appointed by the EC (European Commission 2016) begins by describing EU research as in the European paradox: "When looking ahead to the future of Europe in a globalising world, the contrast is striking between Europe's comparative advantage in producing knowledge and its comparative disadvantage in turning that knowledge into innovation and growth;" the next sentence is a motto "Europe is a global scientific powerhouse" (European Commission 2017b, p. 7). Regretfully, this report is the foundation for the preparation of the Horizon Europe research program for the 2021–2027 period. This program is based on the notion that "Europe is a world leader in science" (European Commission 2018a, p. 11); and that "In a swiftly changing world, Europe's success increasingly depends on its ability to transform excellent scientific results into innovation that have a real beneficial impact" (European Commission 2018b, p. 1). Consistent with these statements, EC optimism is also demonstrated in the press releases; for example: "A new programme – Horizon Europe – will build on the achievements and success of the previous research and innovation programme (Horizon 2020) and keep the EU at the forefront of global research and innovation" (http://europa.eu/rapid/press-release_IP-18-4041_en.htm, accessed on 7 March, 2019). Many examples of this type demonstrate that the EC is not aware that research in the EU is weak. If the EU research policy is based on these convictions, it is likely that the EU research weakness will never be corrected.

It seems evident that the first step in an efficient research policy is that the EC recognizes and addresses the weaknesses of EU research, especially in technology. Only this would make it possible for the EU to return to past research successes. To achieve this, it is necessary to widen the description and characterization of this weakness, expressing the results in terms that are clear to a wide audience.

This purpose, however, is more complex than it would be in a developing country, where simple indicators can be used. The EU is a powerful economy, it was a world scientific leader not so long ago, it still produces an enormous amount of scientific research, and it has competitive industries that are able to maintain a high level of incremental innovations. The current EU research weakness cannot, therefore, be correctly described using simple indicators that are unable to detect complex problems. For example, the statements: "The EU is a global research powerhouse responsible for one-fifth of all R&D investments worldwide" and "In terms of overall scientific production, Europe is in the lead, ahead of the United States and China" (European Commission 2018c, p. 78 and 154) are true, but they are irrelevant truths. Investments and overall scientific production say nothing about discoveries and knowledge advancements, which are the driving force of technological advances and the roots of radical innovations. These achievements define the real research powerhouses and they are where EU research fails.

**2. Aims of this study**

This study aims to further characterize the weakness of the EU research performance described above and investigate its causes. For this purpose it focuses on two research fields of high socioeconomic importance: advanced technology and basic medical research, and compares the performance of a small number of countries within the European Research Area (ERA) with excellent research indicators with that of countries that represent the generally low performance of most EU countries (Brito and Rodríguez-Navarro 2018b; Rodríguez-Navarro and Brito 2018b). As a second tool, this study focuses on the research funded by the European Research Council (ERC), as a means to single out elite and generously funded research in high and low performance countries.

**3. The decline of research in some but not in all European countries**

The most remarkable characteristic of research in Europe is that at the beginning of the twentieth century, Europe (excluding Russia) was not only a global scientific powerhouse (Davies 1997), but the only global scientific powerhouse attending to the number of Nobel prizes (Heinze et al. 2019). In the first 14 years of Nobel Prize awards before World War I (1901–1914), out of a total of 49 Nobel laureates in natural sciences, 46 were Europeans. Comparable figures over a recent 14-year period (2001–2014) are more difficult to tally because of the large number of shared prizes, but it is remarkable that there was not a single year without awards to USA researchers, which was not the case for Europe. This decline mainly affected Germany and France, however, less so the UK (Gros 2018); see also Bauwens et al. 2011). Thus, the most important cause of the decline in European research has been the decline of German and French research, while some northern EU countries have not been affected—or to a lesser extent—by the decline.

Consistent with their scarcity in Nobel awards, the current research performance of Germany and France at the forefront of technological and biological research is disappointing, and well below the global average. This low performance does not occur in all EU countries; in fact, the UK and the Netherlands (in the EU) and Switzerland (in the ERA) maintain highly competitive research (Brito and Rodríguez-Navarro 2018b; Rodríguez-Navarro and Brito 2018b).

These large differences across countries can lead to an incomplete perception of the weakness of EU research if they are not taken into consideration; however, most of the studies that have so far examined the weakness of EU research have considered the EU as a whole (e.g., Albarrán et al. 2010; Dosi et al. 2006; Herranz and Ruiz-Castillo 2013; Rodriguez-Navarro and Narin 2018), and this does not help in finding solutions to a problem that needs to be addressed at the country level. According to population, size, and degree of socioeconomic development, the set of countries with the greatest responsibility for weak research performance in the EU is Germany, France, Italy, and Spain (henceforth GFIS; see Table 2 in Bauwens et al. 2011; Tables 1 and 3 in Brito and Rodríguez-Navarro 2018b; and Table 5 in Rodríguez-Navarro and Brito 2018b). These countries represent approximately 50% of the EU population and 55% of its GNP, and, therefore, the study of these four countries, rather than the study of the whole EU, should give a more clear diagnostic of the difficulties of EU research in competing at the forefront of knowledge.

**4. Research performance can be robustly measured**

Discrepancies between different types of research assessment have their roots in the difficulties of this type of assessment. These difficulties explain why, for over 20 years, EU research policy has been based on the wrong diagnoses of the performance of the EU research system (Dosi et al. 2006). Bibliometric indicators based on counting publications and citations have been used for a long time (Godin 2003, 2006). However, if the purpose is to estimate contribution to the advancement of knowledge then most of these indicators give misleading information. This is because only a very low proportion of the total number of publications reports important scientific breakthroughs. This can be explained by distinguishing between "normal" and "revolutionary" science (Kuhn 1970), and also by taking into consideration that so far "the benefits of scientific discoveries have been heavy-tailed" (Press 2013, p. 822). This implies that the evaluation of the publications in this tail is what really counts in research assessment. However, because the identification and evaluation of these publications is difficult, most indicators "are largely based on what can easily be counted rather on what really counts" (Abramo and D'Angelo 2014, p. 1130), which has been also expressed as "not everything that can be counted counts, and not everything that counts can be counted" (Cameron 1963, p. 13).

Counting the number of Nobel prizes has occasionally been used as a reliable indicator of research success (e.g., Braun et al. 2003; Charlton 2007b, 2007a; Heinze et al. 2019; Schlagberger et al. 2016), but this method has strong limits because Nobel prizes are awarded to extremely infrequent scientific achievements. It can therefore be applied to big science producers—e.g., the USA, the EU, Massachusetts Institute of Technology (MIT), Harvard University, University of Cambridge, etc.—but in most countries and institutions it is a useless procedure because there are no Nobel prizes to be counted. Although Nobel prizes cannot be used as indicators of research performance, however, they can be used to validate other indicators, which demonstrates that many bibliometric indicators do not correlate with the number of Nobel prizes (Rodríguez-Navarro 2011).

Accepting that what really counts in the progress of science are the infrequent scientific breakthroughs that cannot be counted in most countries and institutions, the conclusion is that research indicators cannot be formulated by simply counting something. The alternative is to calculate the probability or expected frequency of the infrequent achievements that boost the progress of knowledge; it has been previously shown that this can be calculated from the frequency of much less cited papers (Rodríguez-Navarro and Brito 2019b). This is because by considering their number of citations, the rank of local papers expressed as a function of their global rank follows a power law (Rodríguez-Navarro and Brito 2018a). As a mathematical consequence, the distribution of local papers in global percentiles attending to their number of citations also follows a power law (Brito and Rodríguez-Navarro 2018a). This power law determines the $e_p$ index, which allows the probability and expected frequency of the infrequent but very highly cited papers that locate in low percentiles of the upper tail to be calculated (e.g., top 0.01%).

The $e_p$ index is a derivative of the exponent of the power law that percentile frequencies obey; i.e., a mathematical parameter that characterizes the distribution of local papers among the global papers. It reveals the *research efficiency* or *breakthrough potential* (Rodríguez-Navarro and Brito 2018b), namely the efficiency of the system in scaling up from less cited papers to highly cited papers. For example, the decrease of the number of papers in the top 1% of most globally cited papers with reference to the number of papers in the top 10% of most globally cited papers. An $e_p$ index of 0.1 indicates that as the percentile decreases, the number of papers in a country or institutions decreases at the same rate as in global publications. Therefore, if the $e_p$ index is lower than 0.1 in a country or institution, the research performance of that country or institution is worse than the global average. Consequently, in countries that are "global scientific powerhouses" the $e_p$ index has to be notably higher than 0.1—excellent research systems have $e_p$ index values of around 0.2.

The probability that a random paper from a given country or institution reaches a top percentile $x$ is calculated by simply raising $e_p$ to a power, as shown by the formula (Rodríguez-Navarro and Brito 2019b):

$$P(x) = e_p^{(2 - \lg x)} \tag{1}$$

As a first approximation, it can be assumed that important breakthrough papers are linked to radical innovations and future technologies, while incremental innovations in present technologies are linked to less cited papers (Dewar and Dutton 1986); therefore, a meaningful research assessment should also provide the probability and expected frequencies of these papers. The $e_p$ index fulfills this condition because probabilities at all citation levels are powers of the $e_p$ index; probabilities at high citation levels—infrequent achievements/high powers of the $e_p$ index—correlates with the number of Nobel prizes (Brito and Rodríguez-Navarro 2018a) and probabilities at low citation levels—more frequent achievements/low powers of the $e_p$ index—correlates with peer reviews (Rodríguez-Navarro and Brito 2019a; Traag and Waltman 2019).

**5. Rationale and design of this study**

Assuming that, in general terms, the weak research performance of the EU has been demonstrated (Albarrán et al. 2010; Bauwens et al. 2011; Bonaccorsi 2007; Bonaccorsi et al. 2017a; Dosi et al. 2006; Herranz and Ruiz-Castillo 2013; Rodriguez-Navarro and Narin 2018; Rodríguez-Navarro and Brito 2018b), the first aim of this study was to characterize it in more detail. An accurate characterization would allow its causes to be studied, as well as avoiding the "wrong diagnosis and misguided policies" (Dosi et al. 2006, p. 1461) that have characterized EU research for many years. For example, the EU research policy of increasing investments to reach 3% GDP (European Commission 2010a, 2018a) might not result in the expected improvement were the main causes of the weak performance not modified by higher investments. In fact, "the way the money is used is probably as critical as the amount of money itself" (Bauwens et al. 2011, p. 20).

As presented in Section 3, previous studies suggest that responsibility for the weak research performance in the EU lies mainly with GFIS; a study of these four countries as a single set has advantages in comparison to the study of independent countries, not only because the conclusions are more representative—GFIS represent around 50% of the EU's population—but, more importantly, because the study of a set of countries allows for a larger base of research publications, making it statistically more robust. According to the $e_p$ index in technological and biotechnological areas, and basic medical research, research performance in Germany is better than in France, Italy, and Spain; however, the differences are small when taking a competitive country, such as Switzerland, as a reference (Brito and Rodríguez-Navarro 2018b; Rodríguez-Navarro and Brito 2018b). GFIS are thus more similar in their weakness, as compared to Switzerland, than different in their degrees of weakness.

The study of a single set of countries with weak research performances has, nonetheless, the inconvenience of lacking a reference of similar countries—in Europe—with competitive research performance. We therefore selected a reference set comprising the UK, the Netherlands, and Switzerland (henceforth UKNCH) for comparison in the analyses. The comparative research performances of UKNCH in fast evolving technologies and basic medical research vary notably (Brito and Rodríguez-Navarro 2018b; Rodríguez-Navarro and Brito 2018b) because university specialization in Europe (Bonaccorsi et al. 2017b) has a strong effect on small countries. Therefore, although UKNCH research performance is taken as a reference in this study, our results are not indicative of the research performances of the individual countries.

We studied domestic papers at the ERA level—i.e., papers authored by at least one GFIS or UKNH researcher plus others from ERA countries external to GFIS and UKNCH, but none from non-ERA countries. Therefore, collaborations of ERA countries external to GFIS and UKNCH could exist in any of the two sets of independent papers.

As already advanced in Section 2, another basic design of our study involves taking advantage of the EU funding programs, especially the ERC program, to select for different levels of excellence. The rationale is simple: the scientific success of a certain country or association of countries depends on a combination of: the ability of its researchers, their funding resources, and research environment—we use this term in a broad sense, including everything from stocks of knowledge to national evaluation methods (Sandström and van-den-Besselaar 2018). By selecting a specific type of funding some of the country differences are eliminated and further insights are possible.

Statistically, citations follow a lognormal distribution (Rodríguez-Navarro and Brito 2018a; Viiu 2018; and references therein) where the $\mu$ and $\sigma$ parameters of the lognormal function increase in parallel with research success. If a subpopulation in a given country is made up of the most capable researchers—e.g., in an elite research university—it is certain that, *ceteris paribus*, the lognormal citation distribution of their publications will have higher $\mu$ and $\sigma$ parameters than that of the total population. Because countries have complex research systems made up of multiple subpopulations of researchers, a low average capacity can be explained by (i) the low performance of all researchers due to a generally poor research environment and insufficient funding, and (ii) the existence of many low research performance institutions that conceal the high performance of a few elite institutions.

These two possibilities can be distinguished by studying the ERC-funded GFIS and UKNCH publications. The ERC was created to promote excellence in science (Celis and Gago 2014; Luukkonen 2014), and it is based on generous funding and the selection of the most excellent research projects. Generous funding could correct insufficient funding in a country or set of countries, but not a poor research environment. It is therefore likely that studying the ERC-funded subpopulations will provide information about the research environment in GFIS and UKNCH.

ERC-funded GFIS research was also compared with the research of an elite institution. The ERC funding of research projects represents an *ex ante* selection that is significantly different from the selection of researchers in elite research institutions, whose members are mainly selected via an *ex post* research assessment improved with multiple considerations of future perspectives. The selected researchers then obtain generous funding. We studied the effect of these two models for selecting excellence by comparing MIT and ERC-funded GFIS publications.

**6. Analyses based on lognormal distributions**

Most results in this study were obtained through the use of the $e_p$ index, which is based on analysis of the distribution of local publications among global publications (Rodríguez-Navarro and Brito 2019b). Citation distributions are lognormal (Rodríguez-Navarro and Brito 2018a; Viiu 2018; and references therein) and local comparisons can also be performed by omitting the comparison with global publications through the more direct but less accurate method of comparing lognormal distributions of citations. In this method, once the parameters of the lognormal distributions of a country, institution, or subpopulation of papers have been calculated, the cumulative probability at any citation level can be calculated. In absolute terms, local probabilities do not have a clear meaning, but this meaning can be obtained through comparison with a reference or gold standard that publishes a similar number of papers. As already mentioned, MIT publications were used as a reference.

The formulas of the lognormal distribution and upper cumulative distribution for a paper to receive more than $C_a$ citations are the following (Aitchison and Brown 1963):

$$p(C, \mu, \sigma) = \frac{1}{\sqrt{2\pi}C\sigma} exp\left[-\frac{(\ln C - \mu)^2}{2\sigma^2}\right] \quad (2)$$

$$p(C_a) = \int_{C_a}^{\infty} p(C, \mu, \sigma) \, dC \quad (3)$$

**7. Technologies selected in this study**

The EC reasonably places emphasis on the importance of research for the economic future of the EU. For example, one of the main features of the new "Horizon Europe" research program is to "foster the EU's industrial competitiveness and its innovation performance, notably supporting market-creating innovation via the European Innovation Council and the European Institute of Innovation and Technology" (European Commission 2018e).

The research areas and topics to be investigated were selected in two independent fields, physical and chemical technologies (henceforth TECH) and biological technologies and basic medical research (henceforth BIO-MED), which are currently at the forefront of knowledge. They were selected in two previous studies (Brito and Rodríguez-Navarro 2018b; Rodríguez-Navarro and Brito 2018b), with a slight increase in the scope in the case of technology (see next section). These research areas and topics will probably continue being important and supporting market-creating innovations and healthcare advancements in the near and medium future.

**8. Methods**

As described above (Section 3), the assessments of this study were based on an analysis of the percentile distribution of citations through the $e_p$ index, after fitting the data to a power law. For this purpose, we counted the number of papers from the two sets of countries and MIT that were in global top percentiles according to the number of citations, as described previously (Brito and Rodríguez-Navarro 2018a; Rodríguez-Navarro and Brito 2018b). In previous cases, we counted the number of papers in six top percentiles from 7 to 35, but for ERC publications we added top percentiles 3 and 1, because, in some cases, the percentile distribution of ERC publications slightly deviates from the power law. In these cases, the $e_p$ index obtained from fitting the power law to the higher percentiles was higher than those obtained from fittings to the lower percentiles. When this occurs, to better predict the probability of breakthrough papers, which locates in top 0.01%, we normally fitted the power law to the lower percentiles instead of fitting the power law to all data points.

This deviation of ERC publications from the characteristic power law that describes the percentile distribution of citations is surprising and suggests that it might be due the peculiar characteristics of ERC selection process (Section 9.5). This intriguing possibility deserves a specific study that is out of the scope of this paper.

Bibliometric searches were performed in the Science Citation Index Expanded of the Web of Science Core Collection (WoS), using the "Advanced Search" feature. For TECH searches we used (TS=(ionic liquid* OR liquid electrolyte* OR liquid salt* OR energy transfer OR fuel cell* OR quantum dot* OR composite material* OR transistor* OR semiconductor OR superconductor OR graphene OR batter* OR solar cell* OR electronic OR metal organic framework* OR nano*) OR SU=Telecommunications)). For BIO-MED searches we used (SU=((biochemistry & molecular biology OR biotechnology & applied biotechnology OR cell biology OR microbiology) NOT (computer science OR mathematical & computational biology)) OR TS=((cancer OR crispr* OR microbiota OR stem cell* OR immunity OR inflamma*) NOT (statistics OR trial OR survey))). For ERC founded publications we used FT=(ERC OR (European Research Council)); for EU funded research excluding ERC and Marie Curie funded publications, we used FT=(((COST OR FEDER OR FP7 OR FP6 OR (European Social Fund) OR (European Regional Development Fund) OR (European Commission)) NOT (ERC OR (European Research Council) OR Marie Curie))). Although Marie Curie (MC) funding was excluded in these searches MC publications were not treated as ERC publications because they have a lower level of excellence (results not shown).

We retrieved only "articles," which excludes review papers, because review papers often receive more citations than the original articles on which they are based. Searches were performed between February 23 and March 5, 2019. Some countries or sets of countries were analyzed on different days but each analysis on a different day was complete, including world and country citation distributions. Because the ERC program started in 2007, our citation analyses were performed for the years 2011–2014.

To calculate the $\mu$ and $\sigma$ parameters of lognormal distributions of citations of MIT and ERC-GFIS publications, we retrieved the number of citations of all publications and used the maximum likelihood method to fit the empirical data to a lognormal function.

## 9. Results

### 9.1. General appraisal of EU research performance

Before going into more detail, we obtained a general appraisal of the research performance in the EU from *Science & Engineering Indicators*, published by the National Science Board of the National Science Foundation (National Science Board 2016, 2018). Among other indicators they report the proportions of papers within five citation-based global percentiles in 13 research areas for years 2002, 2004, 2012, y 2014. Although maximum differences between USA and EU research occur in topics that are at the forefront of technological knowledge (Bonaccorsi 2007; Rodriguez-Navarro and Narin 2018; Sachwald 2015), the National Science Board data enables a general appraisal of research performances in the EU and the USA. Table 1 records the $e_p$ index values in all these 104 cases—8 years and 13 research areas—which clearly indicate that while the USA is above the world average ($e_p \approx 0.12$) the EU is almost exactly at the world average ($e_p \approx 0.10$). Only in the case of "Other Life Sciences" was the EU slightly ahead of, or on par with, the USA.

**Table 1**

Values of the $e_p$ index calculated from the data reported in the *Science and Engineering Indicators*. Data for years 2002 and 2012 in 2016 report, and data for years 2004 and 2014 in 2018 report

| Research field | USA | | | | EU | | | |
|---|---|---|---|---|---|---|---|---|
| | Year | | | | | | | |
| | 2002 | 2004 | 2012 | 2014 | 2002 | 2004 | 2012 | 2014 |
| Engineering | 0.126 | 0.138 | 0.133 | 0.134 | 0.096 | 0.105 | 0.109 | 0.102 |
| Astronomy | 0.127 | 0.119 | 0.125 | 0.132 | 0.102 | 0.109 | 0.109 | 0.107 |
| Chemistry | 0.134 | 0.131 | 0.127 | 0.114 | 0.093 | 0.093 | 0.093 | 0.091 |
| Physics | 0.128 | 0.134 | 0.145 | 0.143 | 0.102 | 0.099 | 0.113 | 0.116 |
| Geosciences | 0.118 | 0.119 | 0.131 | 0.130 | 0.095 | 0.103 | 0.109 | 0.111 |
| Mathematics | 0.121 | 0.126 | 0.109 | 0.112 | 0.091 | 0.096 | 0.095 | 0.106 |
| Computer sciences | 0.137 | 0.139 | 0.142 | 0.138 | 0.082 | 0.082 | 0.100 | 0.090 |
| Agricultural sciences | 0.129 | 0.118 | 0.127 | 0.128 | 0.100 | 0.106 | 0.118 | 0.112 |
| Biological sciences | 0.119 | 0.119 | 0.129 | 0.132 | 0.095 | 0.100 | 0.113 | 0.111 |
| Medical sciences | 0.127 | 0.126 | 0.132 | 0.135 | 0.096 | 0.100 | 0.117 | 0.118 |
| Other life sciences | 0.103 | 0.102 | 0.108 | 0.102 | 0.104 | 0.104 | 0.112 | 0.111 |
| Psychology | 0.115 | 0.111 | 0.121 | 0.110 | 0.084 | 0.102 | 0.096 | 0.105 |
| Social sciences | 0.120 | 0.119 | 0.118 | 0.119 | 0.083 | 0.094 | 0.100 | 0.107 |
| | | | | | | | | |
| Mean | 0.123 | 0.123 | 0.127 | 0.125 | 0.094 | 0.100 | 0.106 | 0.107 |

*Science and Engineering Indicators* 2016 and 2018, National Science Board, National Science Fundation, Arlington, VA. Appendix Tables 5-59 and 5-48, respectively. In the $e_p$ index, the higher the better.

In "Engineering," the means of the $e_p$ index values of the four evaluations is 0.13 for the USA and 0.10 for the EU. The top 0.01% of most cited papers reasonably represents the percentile where most breakthrough and landmark publications concentrate (Bornmann et al. 2018; Brito and Rodríguez-Navarro 2018a), and the probability that a random paper locates in this percentile—$e_p$ index raised to the fourth power, Eq. (1)—reveals the efficiency of research systems in making discoveries. The probability that a random paper reaches the 0.01 percentile in the field of "Engineering" is 2.9E-04 in the USA and 1.0E-04 in the EU, 2.9 times higher in the USA than in the EU. This is a notable difference that implies that the EU should publish 2.9 times more papers to obtain the same number of achievements.

The *Science & Engineering Indicators* reports have no data that allows a comparison of GFIS and UKNCH research. Therefore, we obtained a general overview of their differences by counting the number of universities among the top universities in the Leiden Ranking (Ranking 2019). For this purpose, we counted the number of GFIS and UKNCH universities among the top 25, 50 and 100 universities ordered by the $P_{top\,1\%}$ in the field of "Physical sciences and engineering," in two time periods, 2006–2009 and 2014–2017 (Table 2). The differences between UKNCH and GFIS were striking in the top 25, because in the two periods UKNCH had four and five universities while GFIS had none. In the top 50, UKNCH has eight and seven, and GFIS has three and zero universities in the two periods, respectively. In the top 100, again UKNCH performed better than GFIS. In the first period GFIS has 10 universities that decreased to only four in the second period. Differences between the two periods were due to the emergence of China as a global scientific powerhouse, which impaired the rank of GFIS universities the most, and had a lesser effect on UKNCH than on the USA universities.

**Table 2**

Number of universities among the top 25, 50 and 100 in the CWTS Leiden Ranking in periods 2006–2009 and 2014–2017 in the research field of "Physical sciences and engineering," $P_{top\,1\%}$ indicator

| Country/ Set | 2006–2009 | | | 2014–2017 | | |
|---|---|---|---|---|---|---|
| | Top 25 | Top 50 | Top 100 | Top 25 | TOP 50 | Top 100 |
| USA | 17 | 28 | 45 | 11 | 20 | 31 |
| UKNCH | 4 | 8 | 17 | 5 | 7 | 11 |
| GFIS | 0 | 3 | 10 | 0 | 0 | 4 |
| China | 1 | 3 | 7 | 7 | 17 | 35 |

UKNCH: the UK, Netherlands, and Switzerland; GFIS: Germany, France, Italy, and Spain. Source https://www.leidenranking.com/ranking/2019/list, accessed May 22, 2019

*9.2. EU-funded publications in TECH and BIO-MED research*

This study focuses on TECH and BIO-MED; the number of papers published by GFIS and UKNCH in BIO-MED is rather high. In BIO-MED, it represents about 17-18% of the total number of publications; in TECH the proportion decreases to approximately 13% and 10% in GFIS and UKNCH, respectively (Table 3). These figures show that in addition to its economic importance, we are dealing with a paramount research activity in the investigated countries. The difference in the total number of papers between UKNCH and GFIS roughly reflects their population difference.

**Table 3**
Total number of publications, and number of publications in chemical and physical technologies (TECH), and biotechnology and basic medical research (BIO-MED) from Germany, France, Italy, and Spain (GFIS), and from UK, Netherlands, and Switzerland (UKNCH)

| Year | GFIS | | | UKNCH | | |
|---|---|---|---|---|---|---|
| | Total | TECH | BIO-MED | Total | TECH | BIO-MED |
| 2011 | 143,017 | 17,801 | 24,994 | 62,048 | 5,606 | 11,171 |
| 2014 | 149,224 | 20,080 | 26,828 | 63,066 | 6,250 | 10,884 |
| 2018 | 144,404 | 19,806 | 24,453 | 61,590 | 6,533 | 10,030 |

The data in Table 3 shows a situation of concern for both UKNCH and GFIS because the number of publications did not increase in the period from 2011 to 2018, when it is well known that several Asian countries were contributing to a notable increase in the number of publications and research advancements in these technological areas (Brito and Rodríguez-Navarro 2018b; Rodriguez-Navarro and Narin 2018; Rodríguez-Navarro and Brito 2018b). The consequence is that the share of global publications in TECH and BIO-MED have been continuously decreasing from 2011 to 2018 in both GFIS and UKNCH.

This decrease in the share of publications is especially worrying for GFIS because the $e_p$ index values for TECH and BIO-MED publications were (0.06–0.07) much lower than the world average (Table 4). Taking these two facts together, the effect is that the research by GFIS in technological areas and basic medicine could eventually become of low global relevance in terms of scientific advancement. As described in Section 5, our next step was thus to study the reasons for this low research performance by analyzing the subpopulations of elite papers selected according to their funding sources.

The subpopulation of papers selected by their EU funding in programs other than ERC and MC was higher in GFIS than in UKNCH in the two technologies (Table 4; 14% versus 12% in TECH, and 8.6% versus 5.4% in BIO-MED, for GFIS and UKNCH respectively). In GFIS the selection by this EU funding involved a weak increase of the $e_p$ index, a roughly 10%, with reference to that of the total number of papers. This indicates a practically nonexistent selection of above-average research in these programs. In contrast, this funding clearly increased the $e_p$ index in UKNCH, from 0.10 to 0.14, which indicates a notable selection of above-average research.

**Table 4**
Publications in TECH and BIO-MED from Germany, France, Italy, and Spain (GFIS) and from the UK, the Netherlands, and Switzerland (UKNCH): number and proportion of publications, and $e_p$ index by type of funding. Year 2014

| Country set/funding | TECH publications | | | BIO-MED publications | | |
|---|---|---|---|---|---|---|
| | Number | Percent | $e_p$ | Number | Percent | $e_p$ |
| **GFIS** | | | | | | |
| All papers published | 20080 | 100 | 0.062 | 26828 | 100 | 0.066 |
| EU funded not (ERC or MC) | 2813 | 14.0 | 0.066 | 2305 | 8.6 | 0.075 |
| ERC funded | 726 | 3.6 | 0.143 | 510 | 1.9 | 0.166 |
| **UKNS** | | | | | | |
| All papers published | 6250 | 100 | 0.104 | 10884 | 100 | 0.103 |
| EU funded not (ERC or MC) | 751 | 12.0 | 0.145 | 592 | 5.4 | 0.144 |
| ERC funded | 534 | 8.5 | 0.191 | 414 | 3.8 | 0.271 |

The study of the ERC funded papers is more interesting because ERC funding is based on the selection of the highest research excellence through high quality peer review (Celis and Gago 2014; Luukkonen 2014). The proportion of this elite subpopulation (Table 4) was considerably lower in GFIS than in UKNCH—3.6% versus 8.5% and 1.9% versus 3.8% in TECH and BIO-MED, respectively—which once more denotes the lower research competitiveness of GFIS. Interestingly, despite the apparently more stringent selection by ERC funding in GFIS than in UKNCH, the $e_p$ index had lower values for ERC-GFIS than for ERC-UKNCH papers, 0.14 and 0.17 versus 0.19 and 0.27 in TECH and BIO-MED, and GFIS and UKNCH, respectively.

This lower success of GFIS versus UKNCH in ERC-funded publications in TECH and BIO-MED is a general trend also seen in the number of ERC grants in all fields over the years (Fig. 1). This number increased continuously from 2007 to 2012, when it reached a plateau. Similarly, the number of ERC publications in TECH and BIO-MED increased from 2012 to 2016 and has remained almost constant since then (Fig. 2). Interpretation of these data needs to consider that the size of the GFIS research system is more than twice that of the UKNCH system (Table 3).

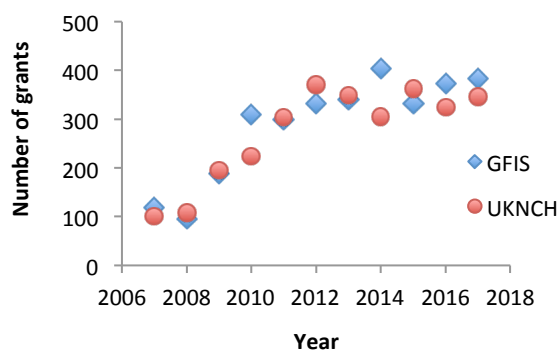

**Fig. 1**. Number of grants from the European Research Council (ERC) awarded to Germany, France, Italy, and Spain (GFIS) and to the UK, the Netherlands, and Switzerland (UKNCH) from 2007 to 2017. Source: https://erc.europa.eu/, accessed 9 April 2019; Proof of Concept grants were not included

*9.3. Evolution of the ERC-GFIS, ERC-UKNCH, and MIT publications*

As already argued (Section 5), low $e_p$ index values in a given country might arise in two different research scenarios: publications were produced by either a homogeneous population of researchers of low competitiveness—most probably because of a poor research environment—or by a heterogeneous population of researchers, where some were highly competitive but others were less competitive. These two cases may be distinguished through the study of the ERC-funded publications because ERC funding implies a selection for excellence. For better representativeness, the study must include several years, because the $e_p$ index shows annual variations in the hot topics of this study.

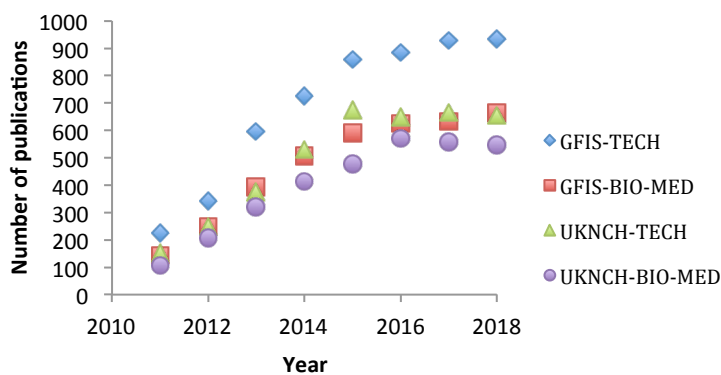

**Fig. 2**. Number of ERC publications in rapid evolving physical and chemical technologies (TECH) and in biotechnological and basic medical research (BIO-MED) from Germany, France, Italy, and Spain (GFIS), and from the UK, the Netherlands, and Switzerland (UKNCH) in years 2011–2018

Figure 3 shows the $e_p$ index values for 2011–2014 for ERC-GFIS, ERC-UKNCH, and MIT publications in TECH and BIO-MED. In TECH there was no difference between ERC-UNKS and MIT publications, while ERC-GFIS publications showed lower $e_p$ index values. The $e_p$ index in ERC-UKNCH and MIT publications varies around 0.2, while in ERC-GFIS publications it was around 0.15; the $e_p$ index values for the three actors can be considered constant from 2011 to 2014. In BIO-MED, the $e_p$ index values in ERC-UKNCH publications varies from 0.15 to 0.28, always below those of MIT, which varies from 0.23 to 0.33. In the case of GFIS the $e_p$ index values varies around 0.15. In both ERC-UKNCH and MIT publications, there was a clear increase in the $e_p$ index from 2011 to 2014, while it remained constant in the case of ERC-GFIS publications. Although irrelevant regarding the $e_p$ index, it is worth noting that the number of ERC-funded publications was continuously increasing from 2011 to 2014 (Fig. 2) while the MIT papers remained constant around 550 (results not shown).

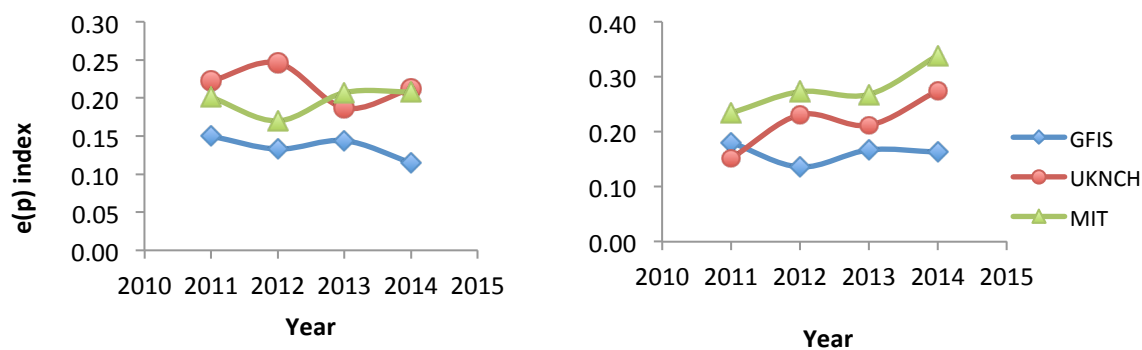

**Fig. 3**. Values of the e(p) index in years 2011–2014 of the ERC funded publications from Germany, France, Italy, and Spain (GFIS) and the UK, the Netherlands, and Switzerland (UKNCH); and publications from the Massachusetts Institute of Technology (MIT). Physical and chemical technologies (TECH; left panel) and biotechnology and basic medical research (BIO-MED; right panel). Curves are drawn to guide the eye.

*9.4. Calculations based on the lognormal distributions: ERC-GFIS versus MIT publications*

A way of comparing the research performances of two countries or institutions is by comparing the probabilities of publishing very highly cited papers. These probabilities can be calculated from the $e_p$ index values Eq. (1) and from the their lognormal citation distributions Eq. (3). These two methods can be made equivalent, without any reference to the global papers, by selecting a number of citations for Eq. (3) that reasonably represent the highly cited tail that corresponds to the percentile used in Eq. (1). We used 1000 citations for the papers published in 2011 and reduced the numbers in proportion to the lower citation window for the papers published in 2012–2014.

Table 5 summarizes the MIT and ERC-GFIS probabilities and their ratios calculated following both approaches—Eq. (1) and Eq. (3). The results could not be equal because there was not an exact equivalence between the 0.01 percentile applying the $e_p$ index method and the number of citations reasonably but arbitrarily selected for the lognormal calculation. Nonetheless, results are totally consistent and exhibited similar annual variations; they unequivocally show that ERC-GFIS research competes poorly with MIT research in TECH and BIO-MED.

**Table 5**
Probability that a random ERC-GFIS or MIT paper is highly cited, calculated from the lognormal distribution of citations and from the $e_p$ index; and MIT/ERC-GFIS probability ratio. The top 0.01 percentile was fixed for the $e_p$ index method; for lognormal calculations, the number of citations was arbitrarily selected as described in the text

| Year | Lognormal ERC-GFIS | | Lognormal MIT | | Lognormal probability | | | $e_p$ probability top 0.01 percentile | | MIT/ERC-GFIS Prob. ratio | |
|---|---|---|---|---|---|---|---|---|---|---|---|
| | $\mu$ | $\sigma$ | $\mu$ | $\sigma$ | Citations | ERC-GFIS | MIT | ERC-GFIS | MIT | $e_p$ | Lognor. |
| | | | | | *TECH* | | | | | | |
| 2011 | 3.458 | 1.196 | 3.420 | 1.339 | 1000 | 0.00195 | 0.00461 | 0.00051 | 0.00163 | 3.2 | 2.4 |
| 2012 | 3.240 | 1.118 | 3.412 | 1.191 | 850 | 0.00086 | 0.00257 | 0.00031 | 0.00084 | 2.7 | 3.0 |
| 2013 | 3.138 | 1.129 | 3.250 | 1.203 | 700 | 0.00126 | 0.00304 | 0.00043 | 0.00184 | 4.3 | 2.4 |
| 2014 | 2.922 | 1.062 | 3.028 | 1.196 | 500 | 0.00097 | 0.00385 | 0.00017 | 0.00187 | 10.7 | 4.0 |
| | | | | | *BIO-MED* | | | | | | |
| 2011 | 3.755 | 1.030 | 3.786 | 1.248 | 1000 | 0.00110 | 0.00617 | 0.00107 | 0.00300 | 2.8 | 5.6 |
| 2012 | 3.398 | 0.934 | 3.592 | 1.212 | 850 | 0.00017 | 0.00463 | 0.00034 | 0.00555 | 16.2 | 27.1 |
| 2013 | 3.325 | 1.068 | 3.566 | 1.290 | 700 | 0.00126 | 0.01036 | 0.00078 | 0.00516 | 6.6 | 8.3 |
| 2014 | 3.081 | 0.954 | 3.491 | 1.262 | 500 | 0.00051 | 0.01543 | 0.00071 | 0.01321 | 18.7 | 30.1 |

## 10. Discussion

*10.1. Wrong diagnoses and misguided policies*

This was the title of the last section and the conclusion of a paper about European research that was written thirteen years ago (Dosi et al. 2006). There have been some changes since then, especially regarding basic research and the creation of the ERC, but essentially the situation has not altered. The EC brags about excellent research in opposition to academic studies (Section 1), but such excellence is an illusion based on incorrect assessments.

As a first approach to further demonstrate the weaknesses of European research, we took advantage of the large amount of data provided by two successive reports of the *Science & Engineering Indicators* (National Science Board 2016, 2018). The advantage of using this data is that they have been generated by the USA's National Science Board and not by EU academics, whose results have never been taken into consideration by the EC. The data recorded by the *Science & Engineering Indicators* include the distribution of papers in 13 research areas and five percentiles for four years across a span of 13 years from 2002 to 2014.

As discussed above (Section 9.1), according to the $e_p$ index values (Table 1), the advantage of the USA over the EU is high. Only in the case of "Other Life Sciences" was the EU slightly ahead of, or on par with, the USA. The top 0.01% of most cited papers reasonably represents the percentile where most breakthrough and landmark publications concentrate (Bornmann et al. 2018; Brito and Rodríguez-Navarro 2018a). In "Engineering," the probability that a random paper locates in this percentile is 2.9 times higher for a USA paper than for a EU paper.

It is worth highlighting that the engineering research field mixes very different technologies and that the 2.9 times difference is just the tip of the iceberg. In fast evolving technology topics, in the EU excluding the UK, the difference can be up to eight times (Rodríguez-Navarro and Brito 2018b). The question raised by these results is whether the EU excluding the UK is prepared to invest eight times more in research than the USA to obtain a similar number of scientific advancements.

In summary, academic publications and very simple analyses (Table 1) show that the EU is not a global scientific powerhouse. If current research policy continues to be based on the same illusory, wrong diagnoses of scientific excellence that have oriented the EU research policy for more than 20 years, then EU research will not correct its weak position. Even worse, the EU had only two relevant competitors 20 years ago, the USA and Japan, but China (together with other Asian countries) is currently in the competition and becoming even stronger than the USA and Japan.

The EU's sustained praise of its research excellence based on wrong diagnoses gives rise to misguided research policies and to a continuous failure in reaching predicted targets. For example, the objective of the Lisbon Strategy launched in 2000, that the EU becomes "the most competitive knowledge based economy in the world by 2010" (European Commission 2010b, p. 2), was an illusory dream unconnected to reality. As it could have been expected, this success was not achieved in 2010 and will not be achieved in 2020. This resounding failure should be a lesson that those responsible for the EU research policy should learn.

*10.2. Research in the EU is heterogeneous*

The EU is not a homogeneous set of countries as regards research (Bauwens et al. 2011; Leydesdorff et al. 2014); some countries are more efficient and some less efficient than the global average. As shown in previous studies (Bauwens et al. 2011; Brito and Rodríguez-Navarro 2018b; Rodríguez-Navarro and Brito 2018b) UKNCH represent the efficient countries while GFIS represent the less efficient countries; according to empirical evidence, the addition to GFIS of Poland, Romania, Greece, Czech Republic, Portugal, Hungary, Bulgaria, Serbia, or a few other countries would increase the size of the set without improving the research efficiency of GFIS. A comparison of GFIS and UKNCH thus provides a reliable picture of the heterogeneity of EU research and of its weakness in most of the EU countries. It has to be taken into account for this comparison that the GFIS research system is more than twice that of UKNCH (Table 2).

The reports of the USA's National Science Board, which we used as a first approach to characterize EU research, do not include data that allows a comparison of GFIS and UKNCH research, but an overview of their differences can be obtained by counting the number of universities among the top global universities with the highest $P_{top\ 1\%}$ indicators in the Leiden Ranking (Table 2). The results show unequivocally the low research level of GFIS with reference to UKNCH and the USA. If GFIS had a UKNCH level, the EU would probably be ahead of the USA in research.

*10.3. Research in TECH and BIO-MED*

Although research in many different fields is important for the life of humans, the knowledge-based economy includes a more limited number of research fields, TECH and BIO-MED are representative samples of these fields. As mentioned above, the research performance in GFIS reasonably represents the average research performance of the whole EU. The low values of the $e_p$ index of GFIS in these specific fields ($\approx 0.06$; Table 4) therefore reveals the that the research performance of the EU in the areas that support the knowledge-based economy is worse than in wide scientific areas ($e_p$ index $\approx 0.1$; Table 1).

This scenario is absolutely baffling. It is startling that some of the EU's weakest research is precisely in the research fields that are crucial for the economy. Many EC documents praise EU achievements in many different fields, such as cancer treatments, solar jet fuel, never ending batteries, exploring the universe, and so on (European Commission 2018d). These achievements are absolutely real; we have already noted that EU research is the research of a powerful economy (Section 1). The important question is therefore whether the number of such achievements is in tune with the size of the EU economy. Assuming that the USA and the EU are of similar sizes in terms of GDP, all the data indicates that the number of important EU achievements in technological fields is potentially six to eight times lower than it should be.

In summary, it is true that the EU is "responsible for one-fifth of all R&D investment worldwide (European Commission 2018c, p. 78), and that it produces a certain number of notable achievements in technology and medicine (European Commission 2018d), but the number of those achievements in the EU is lower than in the USA and probably insufficient to maintain a competitive knowledge-based economy.

*10.4. Research assessments by elite samples of publications in TECH and BIO-MED*

To further investigate the differences between GFIS and UKNCH in TECH and BIO-MED we focused on an elite of research publications.

In Section 5 we considered that there are two potential reasons for the low GFIS values of the $e_p$ index in TECH and BIO-MED (Table 4): (i) a generally weak performance due to a poor research environment, and (ii) a small, highly competitive population of researchers mixed with a large lesser competitive population; these two causes cannot be distinguished by studying the whole production in these research areas. Therefore, we focused on the elite sample of ERC publications. ERC grantees are selected through a rigorous peer-review process that considers both the excellence of the project and the previous scientific success of the applicants (Celis and Gago 2014; Luukkonen 2014); for the EC, the ERC, "in just a few years, has become the point of reference for excellent frontier research in Europe" (European Commission 2017a, p. 27). Our data indicates that there are excellent ERC publications, but their frequency is not the same in all EU countries. The first difference we noticed between GFIS and UKNCH in the two research areas under study, TECH and BIO-MED, was in the proportion of ERC versus the total number of publications. This proportion in GFIS (3.6% in TECH and 1.9% in BIO-MED) was lower than in UKNCH (8.5% in TECH; 3.8% in BIO-MED), which suggested a higher success of UKNCH in obtaining ERC grants.

The inference that a lower proportion of ERC publications in TECH and BIO-MED reveals less success in obtaining ERC grants cannot be checked in terms of grants in these research areas, because grants in these specific areas cannot be distinguished and counted. However, in the whole ERC program, different levels of success can be checked by comparing the total numbers of ERC grants. Although the annual number of ERC grants awarded to GFIS and UKNCH increased more than threefold in the 2007–2018 period, the ratio between GFIS and UKNCH grants remained almost constant over this period (Fig. 1). Considering that the size of the GFIS research system is twice that of UKNCH (Table 3), the success of GFIS in terms of ERC grants is about half that of UKNCH.

The lower success of GFIS in terms of ERC grants may be due to either a lower proportion of applications or a higher number of rejections. We cannot distinguish between these two possibilities with our data, but the former seems unlikely because in less stringent EU-funded programs GFIS is more successful than UKNCH. The ratio between GFIS and UKNCH publications in EU-funded publications in TECH and BIO-MED, excluding ERC and MC publications, is thus around 3.8 (Table 4), which is higher than the GFIS/UKNCH ratio considering the total number of publications (3.2 in TECH and 2.4 in BIO-MED). In contrast, the equivalent ratio in ERC-GFIS/ERC-UKNCH publications is 1.3. In other words, in the less stringent EU funding programs GFIS are more successful than UKNCH and in the stringent ERC funding program is just the opposite. It is unlikely, therefore, that the lower success of GFIS in the ERC funding program is due to less interest; the higher success of UKNCH versus GFIS in the ERC program probably reflects the higher research competence of UKNCH.

Figure 3 shows the values of the $e_p$ index throughout four years, further demonstrating that the excellence of ERC publications is lower in GFIS than in UKNCH—the means of the $e_p$ index value in GFIS, 0.14 in TECH and 0.16 in BIO-MED, were lower than in UKNCH, 0.22 in both TECH and BIO-MED. This is an important conclusion because ERC publications should be at the same level in GFIS and UKNCH—we assume that the selection procedure of grantees is the same for all countries. In other words, the excellence of grantees and projects are similar in GFIS and UKNCH, but the execution of the projects is less successful in GFIS than in UKNCH. We also used MIT publications as an external standard for the comparison of ERC publications (Fig. 3), and the superiority of MIT publications over ERC-GFIS publications is beyond doubt.

Taken together, in terms of breakthrough frequencies, our results question the high level of excellence of the ERC-funded GFIS research. As indicated above, the probability of achieving important breakthroughs equals the $e_p$ index raised to the fourth power, and a similar probability can also be calculated from the lognormal distributions of citations by fixing a certain high level of citations (Section 9.4). Global publications are ignored in the latter approach, which implies that the two calculations are independent—although they are conceptually equivalent and mathematically dependent (Rodríguez-Navarro and Brito 2019b). Table 5 shows that both methods lead to the same conclusion, that of a limited research excellence of ERC-GFIS publications; the coherence of the results strongly supports that there is no flaw in the methods employed.

In summary, GFIS shows lower competence than UKNCH at obtaining ERC grants and GFIS-ERC publications have a lower probability than UKNCH-ERC and MIT publications of reporting breakthroughs.

*10.5. Research environment conditions research performance*

Both MIT and ERC publications are elite samples from the total number of publications in TECH and BIO-MED, but originated from two completely different procedures for selecting researchers. MIT and all other elite research institution attract the brightest researchers because these institutions offer a superb research environment. Once in the institution, these researchers can freely apply for competitive research funding without any specific internal requirements.

The process is completely different in the case of ERC funding. Any researcher from any ERA institution can apply. Only the past scientific performance of the applicant and the content of the project are considered for the selection. No GFIS university is among the top 25 in the CWTS Leiden ranking and there are few among the top 100 (Table 2), which implies that a certain number of ERC grantees can be in universities that do not provide a research environment that is at the expected ERC level. In many of these cases, ERC grantees will have to attend to many hours of teaching and bureaucracy and, at the same time, grantees and hired postdocs will be under great publish-or-perish pressure, which increases the quantity but not the quality of publications. In this research environment the productivity at the forefront of knowledge will be lower for ERC-GFIS grantees than for the MIT researchers. In our opinion this explains why ERC-GFIS publications have a lower likelihood of reporting a scientific breakthrough than MIT publications (Fig. 3; Table 5). Again, the same conclusion is obtained by comparing ERC-GFIS and ERC-UKNCH publications, that equivalent, ERC-funded projects have lower probability of success if executed in GFIS rather than in UKNCH.

The number of ERC-GFIS publications is only 2–4% of the total number of GFIS publications in TECH and BIO-MED (Table 4). If the ERC-GFIS researchers do not compete well at the forefront of knowledge after this stringent selection, it can only be because the whole research system is not competitive and other GFIS researchers will probably be even less competitive. This explains why the $e_p$ index of the publications of the whole population of GFIS researchers in TECH and BIO-MED is only 0.06 (Table 4), well below the world average.

*10.6. Incremental versus radical innovations*

Our study is about research; questions about innovation have to be treated through specific approaches

(as in e.g., Archibugi and Filippetti 2011; Filippetti and Archibugi 2011). However, it is well established that innovation depends on scientific research (Jonkers and Sachwald 2018) and that "although a high level of efficiency can be achieved with incremental innovation performance, radical innovation performance is needed to avoid generating competence traps" (Forés and Camison 2015, p. 831).

It is worth noting that ERC-funded research is addressed to produce breakthrough papers and that many of the top 0.01% of the globally most cited papers report breakthrough achievements; these achievements are basic for radical innovations—innovations that contain a high degree of knowledge. However, a large fraction of all technological progress lies in incremental innovations—innovations that contain a low degree of knowledge (Dewar and Dutton 1986). This type of innovation depends on external absorption and the internal creation of a type of research that might not be very highly cited and that is unlikely to be ERC funded. GFIS, and especially Germany and France, probably compete better in this type of research (Tijssen and Winnink 2018) than at the forefront of knowledge. This possibility, however, should not conceal the undesirable consequences of performing research with low $e_p$ index values. First, because a low $e_p$ index affects the performance of the whole research system and is not linked to any specific citation level, which includes the low-cited papers that support incremental innovations; and second, because present scientific breakthroughs at the forefront of knowledge are the bases for the technology that will be common in the mid-term future. Present-day low numbers of research breakthroughs will thus become future technological dependences.

*10.7. Might the weakness of EU technological research be corrected?*

We have already explained that the illusory dream of the EU becoming the most competitive knowledge-based economy in the world by 2010 was not achieved and might never be achieved. China's scientific production is growing rapidly (Fu and Ho 2013; Leydesdorff et al. 2014) and might reach this target in the near future because China is developing a strong research system (Liu et al. 2017), especially in fast evolving technologies (Kostoff et al. 2007; Rodriguez-Navarro and Narin 2018). Copying China's strategy might be very effective in the EU and it is worth noting that China has been developing its research system *de novo*, which is not the case for many EU countries. The EU's problem with research is thus a problem of research policy. The lower success of Germany or France versus UK in Nobel prizes awards (Gros 2018) or number of highly cited researchers (Bauwens et al. 2011) does not have any other explanation. Should GFIS copy UKNCH research policies then they should achieve a similar success in research.

The low levels of investment of some EU countries in research (European Commission 2010c, p. 9) suggest that the governments of some EU countries are not convinced about the economic benefits of research. This is confirmed in Spain, where the drastic cuts (Pain 2012) that are leading to the dismantling of the research system suggest that Spanish governments see research as a dispensable social activity (Rodriguez-Navarro and Narin 2018). The problem is not only investments, however, because the UK and the Netherlands invest less in research than Germany and France (European Commission 2010c, p. 9), but achieve a much higher probability that a random paper will reach the 0.01 percentile (the data is obtained by dividing the P'$_{top0.01\%}$ indicator by the number of papers, see Table 5 in Rodríguez-Navarro and Brito 2018b).

In some countries, independently of other internal constraints, the repeatedly acclaimed and long-lasting notion of EU research excellence backed by the EC and the *High Level Group* mentioned above (Section 1) might lead to a belief that it is not necessary to increase research. The consequences of this scenario are important because around 85% of EU public investments in research and innovation come from national funding (European Commission 2018a) and the EU "does not wish to usurp national authorities in the management and implementation of these activities"—the subsidiarity principle (European Parliament 2017, p. 10). A necessary first step for the EU in order to return to being a global scientific powerhouse is to recognize its research weakness. In addition, the EC should warn the less competitive countries of their low-efficiency research and eventually penalize this lack of solidarity. Taking the EU's Stability and Growth Pact (https://ec.europa.eu/eurostat/statistics-explained/index.php/Government_finance_statistics, accessed April 30, 2019), as a model, a minimal investment in research could be imposed on all EU country members. In contrast, it would be very difficult for the EC to impose a general research policy.

The reasons for research weakness might be different in each country, because success depends on that several key prerequisites are fulfilled (Bornmann and Marx 2012), which implies that unsuccessful research across countries can have multiple causes. This would explain why two countries that are so different in technological research, Germany and Spain, are similarly distant from Switzerland in terms of the $e_p$ index (Rodríguez-Navarro and Brito 2018b). In Germany, governments are interested in research and fund it generously (Federal Ministry of Education and Research 2018). In contrast, the Spanish government's drastic cuts to research funding (Pain 2012) show that it considers research to be a dispensable activity. The answer to the question of why the German and Spanish $e_p$ index values in technological research are, nonetheless, not too dissimilar might shed light on the general problems of GFIS regarding excellence in research.

A likely possibility is that the difficulties that GFIS researchers experience in carrying out competitive research might be related to their universities (Bauwens et al. 2011). In most advanced countries universities play a decisive role in national research and there might be a causal relationship between weak research and the absence of universities in top positions in the Leiden Ranking. Although this causal relationship has to be specifically studied, it is clear that some types of university governance might be determinant of low research performance (Aghion et al. 2020). For example, a high degree of cronyism and an institutional culture that does not favor competition, creativity, intellectual risk, and openness to the outside world impair research performance (Rodríguez-Navarro 2009). University governances are very different across countries (Paradeise et al. 2009) and general recommendations for improvements (Group 2009) might be of little help. Each country should take care of its own university system, but the results should be controlled by the EC, as in the EU's Stability and Growth Pact. If the EU genuinely wants to recover its past status as global scientific powerhouse and to maintain a competitive, knowledge-based economy in the near future, then strict measures should be taken by all EU countries regarding research. While it is necessary (Pavitt 2000), but not sufficient, to increase research investments, illusory declarations of excellence will certainly impede improvements.

## 11. Conclusions

Our study distinguishes research success between the more competitive (UKNCH) and less competitive (GFIS) countries. It is surprising that among advanced countries as physically close as the Netherlands, Switzerland, Germany, and France, it is the Netherlands and Switzerland that are highly competitive, while Germany and France are much less competitive at the forefront of technological knowledge. The first purpose of the EU research policy should be to correct the deficient research performance observed in GFIS and in many other EU countries. The EU would be a global scientific powerhouse if most of its countries were as efficient as UKNCH are.

The finding that the probability that a random ERC publication reports a breakthrough is higher in UKNCH than in GFIS clearly suggests that the weakness of GFIS research is not just a problem of funding for research projects. Our results support that similarly ERC-funded projects are more successful if they are executed in UKNCH rather than in GFIS. Therefore, although an increase in investment is necessary, the EU will not improve its research performance by exclusively increasing investments and expanding the ERC program. The improvement of the observed low research performance by most of EU country members requires political measures.

In this scenario, the repeated assertion that EU research is excellent operates against the solution of the problem. The conviction of governments that the science performed in their countries is excellent demotivates them from investing more in research and from making reforms that are necessary but unpopular. The failure of the strategic goal of the Lisbon European Council in 2000 of the EU becoming the most competitive knowledge-based economy by 2010 is a lesson that should be learned.

The EU's Stability and Growth Pact only includes financial rules, without considering that economic growth also depends on the generation of knowledge. The future of the EU will not be insured by the sole application of financial rules.

**Acknowledgments.** We thank Juan Imperial and Manuel Pérez-Yruela for insightful and constructive feedback.

**Funding.** Tuis work was supported by the Spanish Ministerio de Economía y Competitividad, Grant Number FIS2017-83709-R.